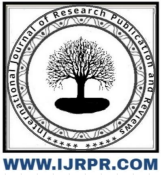

# International Journal of Research Publication and Reviews

# Leveraging Social Media Analytics for Sustainability Trend Detection in Saudi Arabia's Evolving Market

*Kanwal Aalijah\**

*[a] School of Electrical Engineering and Computer Science (SEECS), H-12 Campus, Islamabad, Pakistan*
*[b] National University of Science and Technology (NUST), Islamabad, Pakistan*
*DOI : https://doi.org/10.55248/gengpi.6.0325.1257*

**A B S T R A C T**

Saudi Arabia's rapid economic growth and social evolution under Vision 2030 present a unique opportunity to track emerging trends in real-time. Uncovering trends in real time can open up new avenues for business and investment opportunities. This paper explores how AI and social media analytics can uncover and monitor these trends across sectors like sustainability, construction, food & beverages industry, tourism, technology, and entertainment. This paper focus on use of AI-driven methodology to identify sustainability trends across Saudi Arabia. We processed millions of social media posts, news, blogs in order to understand sustainability trends in the region. The paper presents an AI approach that can help economists, businesses, government to understand sustainability trends and make better decisions around them. This approach offers both sector-specific and cross-sector insights, giving decision-makers a reliable, up-to-date snapshot of Saudi Arabia's market shifts. Beyond Saudi Arabia, this framework also shows potential for adapting to other regions. Overall, our findings highlight how by using AI-methodologies, give decision makers a reliable method to understand how initiatives are perceived and adopted by the public and understand growth of trends.

## 1. Introduction

Social media can be used as a great tool to gauge public opinion and sentiment. Data from social media platforms such as X, Facebook and Instagram can be used to understand how masses engage and evaluate the economic and environmental activities and initiatives. After the announcement of Vision 2030 there was a massive shift in initiatives to reduce the economic dependency on oil. Moreover, as a part of the new economic vision, enhancement in sustainability initiatives become increasingly important.

Vision 2030 proposed many initiatives including sustainability in every aspect of economic development. The vision also proposed the renewable energy projects and climate action projects. Whenever a government proposes a new policy or a vision, it becomes increasingly important to understand what is the sentiment and perception of general public towards the initiative. Many researches, that have been conducted in the past, they put an emphasis on the fact, how important is to evaluate the perception of public towards any initiative [1]. Many studies have demonstrated how effectively sentiment and engagement analysis can be used to evaluate the impact of public policies and agendas [2,3].

The paper presents a novel approach, to analyze and process social media data so that it can be used to understand the public sentiment towards various government initiatives. In this approach, various machine learning models have been used such as sentiment analysis, clustering techniques and predictive modeling. Using these techniques, we have tried to explore how sentiment evolves over time [4]. We also explored which topics in the dataset gain traction and which topics face a decline in the area of sustainability [4,5].

We collected data from various social media websites such as X, Facebook, Instagram, Google News etc. After data collection, data cleaning was performed. Later the posts were filtered using sustainability related topics. The filtered posts were then clustered in order to group them and to understand the discussion that is happening in each of the cluster. The analysis provided a key insight in to public sentiment and helped us understand what are the various factors that drive the success of these initiatives.

## 2. Literature Review

There is an increased growth in the discussions regarding the sustainability initiatives. Researchers are now looking in to the influence of popular trends and topics on the perception of public. Traditionally social media data is used to understand and gauge public sentiment towards trends and issues.

Many studies have previously used social media data to gain insights about public opinion towards topics of interest [6, 7]. Understanding public sentiment plays an important role in understanding how public is perceiving new initiatives in general.

The sentiment analysis has proven as an effective approach to track public sentiment towards various topics. For example, Ahmed and Li presented in their study how sentiment analysis can be used in tracking public responses towards environmental policies [8]. Sentiment trends over time in correlation to other topics are also very beneficial when understanding the public perception about a certain topic. In order to understand the sentiment trends in relation to other topics, many studies have proposed the use machine learning models such as BERT, LSTM, SVM to achieve this goal [9, 10]. Since the launch Vision 2030 in KSA, the researchers have become very interested in understanding the public sentiment towards the initiative [11]

In addition to sentiment analysis, topic modeling techniques such as Latent Dirichlet Allocation (LDA) are also used to perform topic modeling in the datasets to understand the topics of discussion. More recently transformer-based models are also now being used in the analysis of emerging topics and themes [12]. These techniques have also been utilized in the context of sustainability. For example, it can be used to determine public concerns, regarding policy changes, updates etc. This methodology is also used to understand which initiatives resonate most with the public [13, 14]. Once topics are identified they are normally clustered together in order to summarize the discussions. Most used clustering technique is HDBSCAN [15]. This technique is very beneficial in social media data, as we don't have to define the clusters before running the model. The model is unsupervised and defines clusters itself.

In to understand public's interest, we can use engagement metrics. Engagement includes likes, shares, saves and comments. These metrics provide a quantitative measure on how the general public is perceiving the particular sustainability initiatives. Zhang et al. (2019) analyzed social media engagement to assess public support for renewable energy projects [16]. As a result of this study, it was found that high engagement in posts was often linked to transparency in government communications and clear benefits for the community. This highlights the value of engagement analysis in determining which initiatives are gaining public traction and which need further support to foster engagement.

Predictive modeling can be used in forecasting public sentiment trends. Many machine learning models are used in predictive analysis. Models such as Time-series modeling, Recurrent Neural Networks (RNNs), Long Short-Term Memory (LSTM) models etc. are used in predicting future trends [17]. The prediction is based on historical sentiment data. Predictive analysis has given an opportunity to policymakers to understand potential shifts in public opinion by understanding public sentiment. For example, Green et al. (2021) used LSTM models to predict public sentiment regarding climate change initiatives. Early identification of negative sentiment could be used to start targeted awareness campaigns and can be used to improve the general messaging of the initiative [18].

Overall, previous research also emphasizes the importance of monitoring public sentiment, engagement, and trends to understand the public perception. However, there are gaps in the literature regarding the analysis of sustainability-specific topics in the context of Saudi Arabia's Vision 2030. The gap lies in the focus on integrating advanced clustering techniques and predicting future trends using a holistic approach. This study aims to address these gaps. The study proposed a novel approach using NLP techniques, clustering, and predictive modeling to provide a comprehensive analysis of sustainability-related discussions. The findings will contribute to the broader understanding of how public sentiment evolves in response to national sustainability efforts and will provide insights to better align public initiatives with Vision 2030 goals.

# 3. Methodology

In this study we proposed an advanced AI-driven methodology that detects and analyze trends across within the social media data. Additionally, the study also, integrates a predictive formula to identify trends with the highest growth potential. This kind of studies can help business, government and institutions to take proactive decisions focusing on growth trajectories of various trends

## 3.1 Data Collection

The data was collected from major social media platforms, including X, Facebook, Instagram, Tik-tok, Google News, and Reddit. The data was collected for the years 2018 to 2024. Python crawlers were used to crawl the data. They extracted posts description, comments etc. With help of data collected from these platforms, we were able to capture real-time discussions and public sentiment relevant to key sectors [19].

Total number of posts collected was 30 million. This scale ensures adequate coverage across different platforms and provides sufficient data points for sentiment, trend, and network analysis.

## 3.2 Data Preprocessing

a. *Identification of country:* Once the initial data collection was done. We wanted to focus on KSA related posts. Hence, we filtered the data to focus only on content relevant to Saudi Arabia. This filtering was done by checking for posts with geolocation and identifying the specific hashtags. For geo location we checked for Saudi Arabia and for hashtags we checked for #KSA, #Saudi, or city names (e.g., Riyadh, Jeddah, Dammam) mentions in the posts. Posts that did not meet these criteria were considered irrelevant and were discarded. Following this filtering process, we were able to ensure that the dataset is limited to KSA only.

b.   **Removal of Noise:** In this step the text is cleaned to remove URLS, emojis and special characters. These elements can introduce noise and affect the performance of text-based models [20].

c.   **Normalization:** In the process of normalization the text is converted into a consistent format, this addresses the variations in spelling and script. The process is beneficial in handling diverse ways in which Arabic is written on social media [21].

d.   **Tokenization:** After normalization the text was split into individual words or referred to as tokens [22].

e.   **Filtering Stop Words:** Frequently used words are referred to as stop words. These stops words carry little semantic meaning; hence it is better to remove them [23].

f.   **Stemming:** This process involved reducing words to their base or root form. For example, different forms of a word like "writing," "writes," and "written" would be reduced to "write" [24].

g.   **Lemmatization:** The process of lemmatization considers the context and converts the words to their dictionary form. This is an important exercise especially when the data is in Arabic language [25].

h.   **N-gram Generation (Bigrams and Trigrams):** Data on social media is usually of very short length, sometimes it is even less than 10 characters. For this purpose, we generate Bigrams (two-word combinations) and trigrams (three-word combinations). They help us capture the meaning of short social media posts. The n-grams help us understand the common phrases that single words might miss [26].

i.   **Engagement Metrics:** Engagement metrics are likes, comments, saves and shares. We captured them from the posts and maintained them.

j.   **Handling Missing Data:** In order to handle the missing data such as missing engagement metrics or geo-location, we used imputation technique. In this technique we try to retain the majority of the data and information by substituting missing data with a different value. We used the most frequent value in this case and replaced the missing value with it. Imputation techniques basically help in preserving the dataset size and ensures that no significant bias is introduced due to missing data [27].

1.   **Topic Filtering:** After the data was cleaned, we further wanted to target only the posts that align with this study. For this reason, a small dataset was created that contained sustainability related keyword, phrases etc. We filtered the posts using these keywords so that the focus remains specifically on sustainability-related content. Filtering process was done on entire dataset.

2.   **Data Transformation:** After the data was cleaned, we transformed it in to embeddings. Embeddings are numerical representations suitable for modeling. We generated the **text embeddings** using the transformer model **DistilBERT - bert-base-multilingual-uncased-sentiment model** [28]. This helped us to capture semantic information from the posts.

3.   **Sentiment Analysis:** The posts on social media are usually in various languages. Hence, we needed a model that could handle multi lingual data. For this reason, we used DistilBERT [28] as it is designed to handle multi-lingual data. Using this method, we were able to classify the posts into positive, negative, or neutral categories. We fine-tuned the model using a dataset of multilingual social media posts. The data was split into training (80%), validation (10%), and testing (10%) sets. Each posts/tweet/comment/new headline was assigned a sentiment score. This score was then used to understand the overall public sentiment toward sustainability initiatives in Saudi Arabia.

4.   **Topic Clustering:** In order to merge similar topics we performed clustering. This helped identify common themes within the filtered posts. For the process of clustering, Hierarchical Density-Based Spatial Clustering of Applications with Noise (HDBSCAN) [29] was used. This model is very effective in discovering the clusters without requiring the number of clusters to be specified in advance. This kind of approach makes it suitable for social media data. By clustering we were able to merge overlapping discussions, into broader themes that provide a more holistic understanding of sustainability trends in Saudi Arabia.

5.   **Topic Assignment:** Once all the smaller topics/phrases/keywords were grouped in to clusters, each cluster was assigned a topic. This helped us understand the theme of discussion in each topic. The reason of categorizing each post into topic was to perform the further analysis on well-defined topic groups. It helped us with more detailed and targeted exploration of public sentiment and engagement with different sustainability initiatives. This was the final step, now the dataset is ready for the analysis.

*3.3 Analysis*

1.   **Sentiment Trend Analysis:** In order to understand the sentiment trends for each sustainability topic we used **time-series sentiment analysis**. It was visualized that how public sentiment changes towards the sustainability initiatives Many topics were studied in reference to sustainability. The approach allowed us to understand how public sentiment is influenced by various topics [30].

2.   **Topic Engagement Analysis:** We studied engagement metrics such as likes, shares, saves and comments in order to determine which sustainability topics gathered the most interest. In theory, people tend to engage more with the topic of interest [31].

3. **Predictive Analysis:** Various machine learning models can be used to predict future trends. For this study we used LSTM networks [33]. The reason for the selection was that this model is very effective in capturing long-term dependencies in time-series data. This allows us to capture the change in sentiment more accurately.

Using this analysis, we were able to define public sentiment, engagement, and emerging trends related to sustainability in Saudi Arabia. In the next section we will discuss the results of the analysis and connect the dots that how the analysis can help businesses take the right decisions by studying the trends.

# 4. Results

In these sections we present the results of our methodology. We leveraged various machine learning techniques and models to analyze social media data and understand sustainability trends in the region of KSA. We used various models such as HDBSCAN for clustering models, DISTILIBERT for sentiment analysis, and linear regression for trend prediction for upcoming years.

## 4.1 Data Collection and Preparation

The dataset that we collected comprised of 30 million social media posts. The posts were collected from 2018 to 2024. The platforms of interest were X, Instagram, Facebook, Google news. We made a small dataset of sustainability related keywords, that could help us target the right posts. The collected posts were then filtered using the defined dataset.

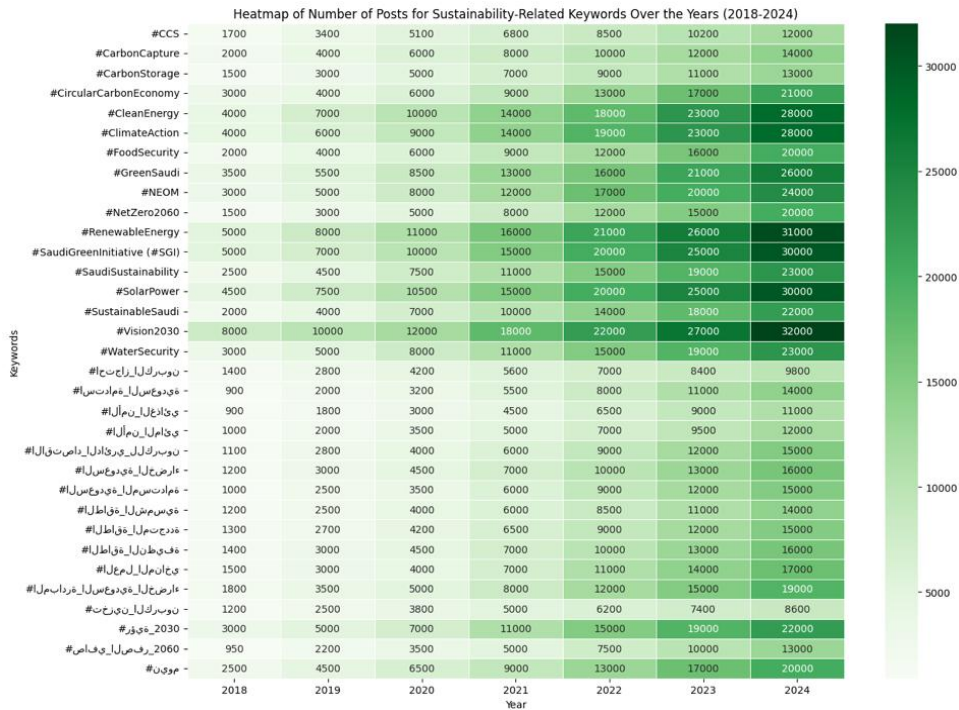

**Fig. 1  Identified Posts for Sustainability Keywords**

The **Fig. 1** shows the frequency of posts for the relevant keywords across the years 2018 to 2024. There is a clear upward trend in the frequency of mentions for most sustainability topics, indicating growing public awareness and engagement with initiatives such as carbon capture (#CarbonCapture, #احتجاز_الكربون), renewable energy (#RenewableEnergy, #الطاقة_المتجددة), and the Saudi Green Initiative (#SaudiGreenInitiative, #المبادرة_السعودية_الخضراء). This figure underscores the increasing emphasis on sustainability in public discourse, supporting the growing focus on environmental goals under Vision 2030.

**Table 1 – Sustainability related posts division per year**

| Year | Total Posts Collected | Posts Related to Sustainability (%) | Sustainability Posts Collected |
|------|----------------------|-------------------------------------|-------------------------------|
| 2018 | 3,000,000 | 8% | 240,000 |
| 2019 | 3,500,000 | 10% | 350,000 |
| 2020 | 4,000,000 | 12% | 480,000 |
| 2021 | 5,000,000 | 15% | 750,000 |
| 2022 | 5,500,000 | 18% | 990,000 |
| 2023 | 4,500,000 | 20% | 900,000 |
| 2024 | 4,500,000 | 22% | 990,000 |
| Total | 30,000,000 | | 4,700,000 |

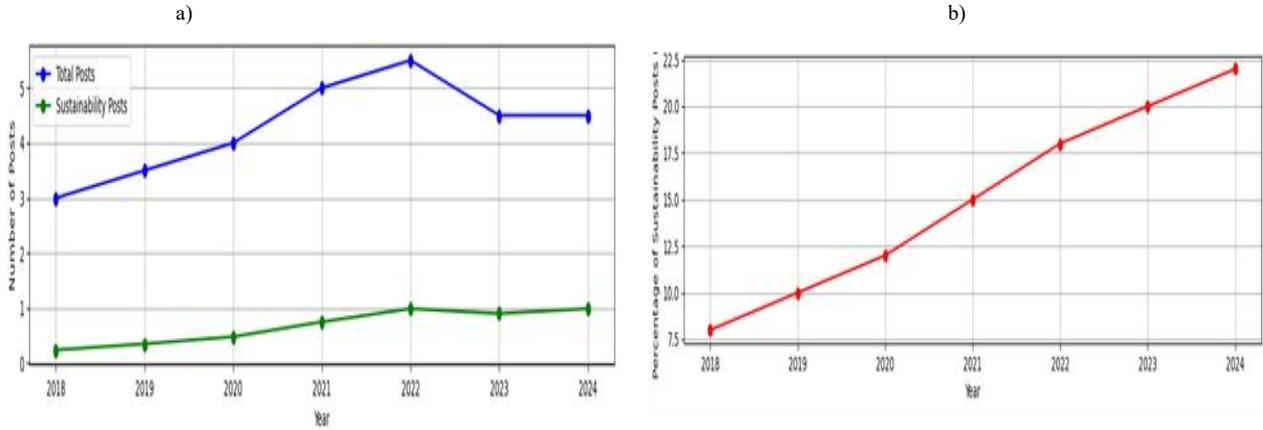

**Fig. 2 - (a) Total Posts Vs Sustainability related posts (b) Percentage of Sustainability related Posts;**

**Fig 2– (a)** shows the comparison of sustainability related posts vs the total collected posts. It can be seen in the **Table 1 and Fig 2– (b)** the percentage of sustainability-related posts increased significantly from **8%** in **2018** to **22%** in **2024**, demonstrating a clear upward trend in public engagement with sustainability topics. This growth indicates a rising interest in sustainability issues among the population, possibly due to the efforts under Vision 2030, which has pushed sustainability initiatives to the forefront of public discourse. The total number of sustainability related posts that our crawlers identified turned out to be 4.7 million, which reflects 15.7% of the entire dataset.

*4.2 Sentiment Analysis on Posts*

It can be seen in **Fig. 3 - (a)** the split of the dataset used for fine-tuning of the sentiment analysis model. Data was basically divided in to training, validation and testing sets in the ration 80:10:10. The approach was used to ensure that model was trainined effectively and was able to handle diverse language variations. On social media, data is generated in various language, this kind of approach is particularly useful for such cases.

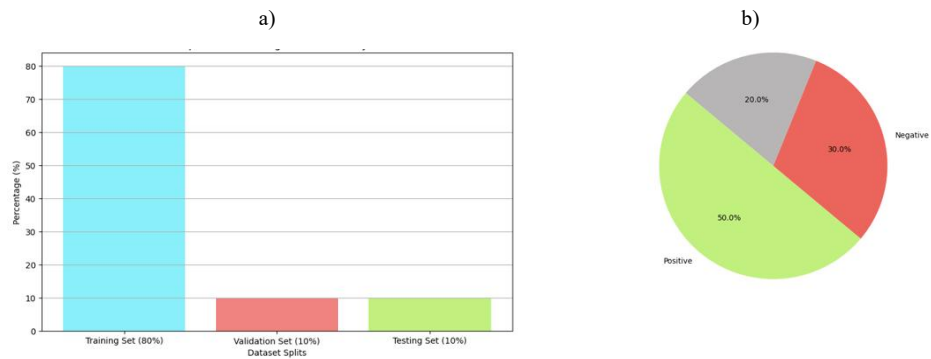

**Fig. 3 - (a) Fine-tuning the sentiment analysis model; (b) Sentiment distribution for sustainability-related posts between 2018 and 2024**

The **Fig. 3 - (b)** illustrates the sentiment distribution for sustainability-related posts between 2018 and 2024. The bert-base-multilingual-uncased-sentiment model classified posts into positive (50%), negative (30%), and neutral (20%) categories. This distribution highlights that a significant portion of posts held a positive sentiment, indicating overall public support for sustainability initiatives in Saudi Arabia.

*4.3 Clustering the topics and assigning them labels*

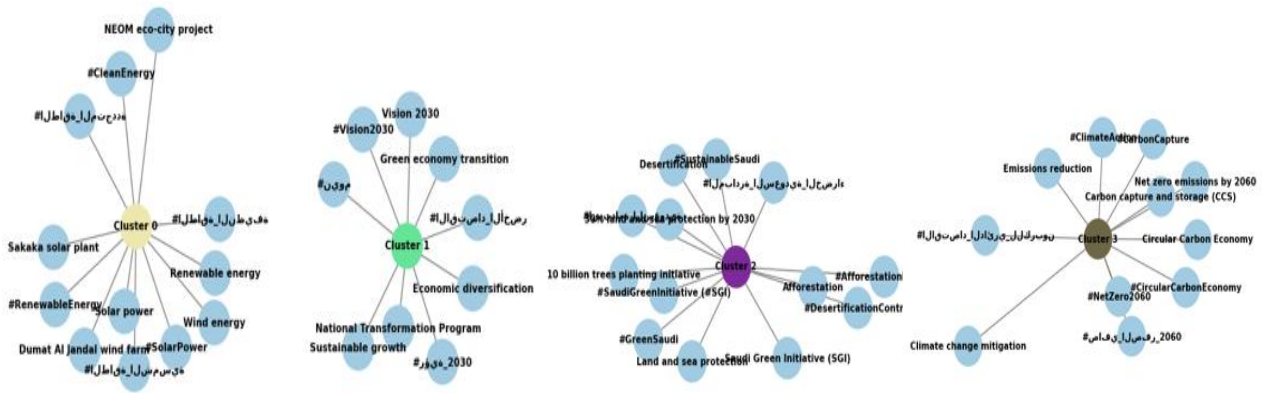

**Fig. 4 – Shows the assignment of keywords, phrases and hashtags to various clusters**

The Fig. 4 shows that the various topics and keywords have been grouped together into the clusters. The model grouped the words together in to 4 final clusters. This process helped us consolidate the information in our dataset. **Table 2** shows the names of each cluster and the keywords that are assigned to the respective cluster. These clusters give a good idea that what kind of discussions are part of each cluster.

**Table 2 – Shows division of hashtags, keywords and phrases in to clusters**

| Cluster ID | Cluster Name | Keywords Assigned |
|---|---|---|
| 0 | Renewable Energy Initiatives | #RenewableEnergy, #CleanEnergy, #SolarPower, Renewable energy, Solar power, Wind energy, Sakaka solar plant, Dumat Al Jandal wind farm, NEOM eco-city project, #الطاقة_المتجددة #الطاقة_النظيفة, #الطاقة_الشمسية |
| 1 | Vision 2030 and Economic Growth | #Vision2030, Vision 2030, National Transformation Program, Economic diversification, Sustainable growth, Green economy transition, #رؤية_2030, #نيوم, #الاقتصاد_الأخضر |
| 2 | Environmental Protection | #SaudiGreenInitiative (#SGI), Saudi Green Initiative (SGI), #GreenSaudi, #SustainableSaudi, Afforestation, #AfforestationProject, #DesertificationControl, #Desertification, #Land and sea protection, #30% land and sea protection by 2030, #10 billion trees planting initiative, #New Afforestation Campaign, #المبادرة_السعودية_الخضراء, #استدامة_السعودية |
| 3 | Climate Action and Carbon Reduction | #ClimateAction, #CircularCarbonEconomy, Circular Carbon Economy, #NetZero2060, #CarbonCapture, Carbon capture and storage (CCS), Emissions reduction, #Climate change mitigation, Climate change mitigation, Net zero emissions by 2060, #الاقتصاد_الدائري_للكربون, #صافي_الصفر_2060 |

*4.4 Trend analysis for the topics*

We performed trend analysis on sentiment timeline as well as on engagement timeline. The reason for this analysis was to understand how the sentiment and engagement grows over time for each cluster. Using predictive model LSTM networks **( Section 3, 3.3)** we made the prediction for upcoming trends for these clusters for the years beyond 2024.

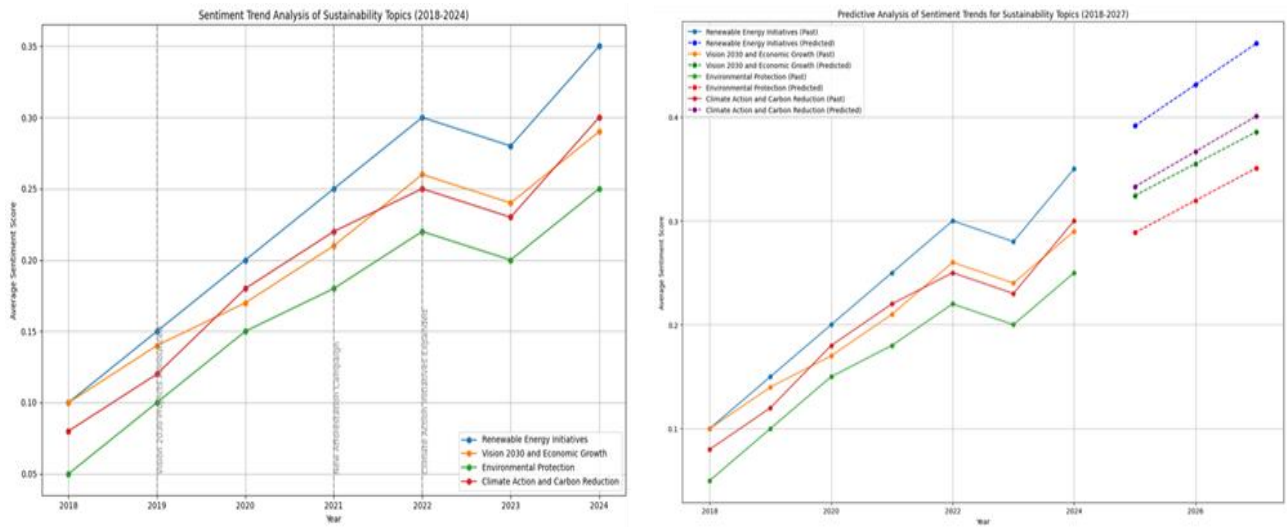

**Fig. 5 - (a) Sentiment Trend Analysis of Sentiment Trends; (b) Predictive Analysis of Sentiment Trend**

Fig 5 (a), presents the sentiment trend analysis for current dataset for key sustainability topics from 2018 to 2024, Fig (b) presents the the predictive trend analysis for the topics for upcoming years. The analysis was conducted using Linear Regression models to predict public sentiment for future years (2025-2027). The sentiment trend lines for each topic—Renewable Energy Initiatives, Vision 2030 and Economic Growth, Environmental Protection, and Climate Action and Carbon Reduction—show both historical data (solid lines) and predicted values (dashed lines). The predictive analysis indicates a generally positive trajectory in public sentiment towards sustainability initiatives, suggesting growing support for renewable energy, climate action, and economic transformation in alignment with Vision 2030.

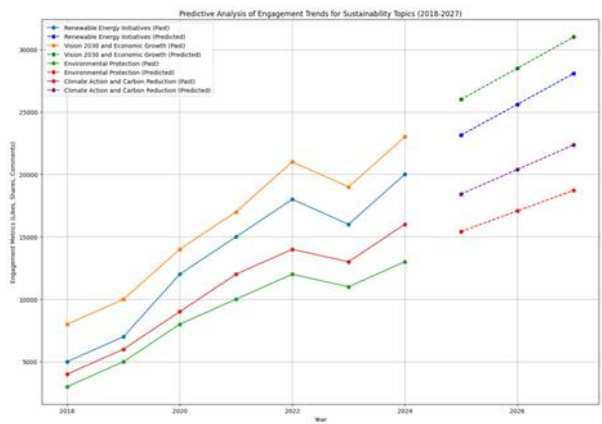

**Fig. 6 - (a) Predictive Analysis of Engagement Trends;**

The engagement trend analysis along with its prediction trend for the upcoming years can be seen in the **Fig 6 (a)**. The dotted lines show the predicted values for these trends for the years 2024 – 2027. This indicates consistent increase in public engagement towards the sustainability topics, reflecting a support towards the topic understudy.

## 5. Conclusion

In this study we presented a novel method, that helps the governments and policy makers understand the public sentiment towards various initiatives and policies. We used various machine learning models such as sentiment analysis models, clustering models, predictive models such as linear regression. We presented structured results that clearly monitor the sustainability trends in the region.

### 5.1 Impact of the study

The results obtained from this study (or similar study performed for a different topic) can be used to understand impact on the sentiment and perception of people towards any initiative. This can be very beneficial for the businesses as they can quickly gauge the public opinion and understand whether the approach they are taking is working in their favor or not. By using a similar approach business can identify various opportunities for investment. In this paper we studied topics like renewable energy, climate action under Vision 2030. Businesses can adjust their strategies, and align them more with the sustainability initiatives to maximum their return on investment. Monitoring these trends enables companies to gain a competitive edge by focusing on sustainability measures that are gaining positive traction, thereby fostering a more positive public image. Monitoring the trends actually enable the businesses gain customer loyalty as well.

### 5.2 Limitations of This Study

The study offers important insights in to sentiment and public perception towards the sustainability initiatives that government is taking under the light of Vision 2030. There are some limitations to this study.

- Since the data is gathered from social media, and there is a chance only a small chunk of might be crawled using social media crawlers. All the analysis that we have performed is on the collected dataset, which might make the results biased and skewed.

- Since machine learning models are used in this study, the results greatly depend upon the accuracy of the models used.

- Currently we are able to scrape all the publicly available data from social media, but if in future the platforms make it difficult to crawl the data, we might have to look in to other datasets.

### 5.3 Future Work of This Study

Future research can expand upon this study by incorporating other sources of public opinion, such as news articles, official statements, and surveys, to provide a more comprehensive understanding of sentiment regarding sustainability initiatives. Advanced modeling techniques, such as deep learning, could be employed to capture the nuances of multilingual sentiment analysis more effectively. Moreover, expanding the scope to compare sustainability sentiment trends across different countries could help benchmark Saudi Arabia's progress within a global context, offering further insights for policymakers and investors.

### Declaration

This study is conducted solely for research purposes and is not affiliated with or intended to promote any business or commercial entity. Its objective is to explore how AI, when applied to social media data, can be used to generate insights into market trends. The findings presented reflect only the data available on social media platforms during the study period and illustrate the types of insights AI methodologies can provide.